# Intrinsic parallel rotation drive by electromagnetic ion temperature gradient turbulence

Shuitao Peng, Lu Wang,* and Yuan Pan

State Key Laboratory of Advanced Electromagnetic Engineering and Technology, School of Electrical and Electronic Engineering, Huazhong University of Science and Technology, Wuhan, Hubei 430074, China

*E-mail: luwang@hust.edu.cn

## Abstract

The quasilinear intrinsic parallel flow drive including parallel residual stress, kinetic stress, cross Maxwell stress and parallel turbulent acceleration by electromagnetic ion temperature gradient (ITG) turbulence is calculated analytically using electromagnetic gyrokinetic theory. Both the kinetic stress and cross Maxwell stress also enter the mean parallel flow velocity equation via their divergence, as for the usual residual stress. The turbulent acceleration driven by ion pressure gradient along the total magnetic field (including equilibrium magnetic field and fluctuating radial magnetic field) cannot be written as a divergence of stress, and so should be treated as a local source/sink. All these terms can provide intrinsic parallel rotation drive. Electromagnetic effects reduce the non-resonant electrostatic stress force and even reverse it, but enhance the resonant stress force. Both the non-resonant and resonant turbulent acceleration terms are also enhanced by electromagnetic effects. The possible implications of our results for experimental observations are discussed.

## 1. Introduction

The importance of plasma rotation for tokamak has been widely discussed. It is well known that plasma rotation can improve tokamak performance via stabilizing macroscopic MHD instabilities such as RWMs [1-4] and NTM [5, 6] and suppressing micro-turbulence [7-9]. Neutral beam injection (NBI) is mainly used to drive plasma rotation in most of current tokamaks, but it is of limited utility and high cost in future ITER and reactor-grade tokamaks which have large size and high density plasmas. Therefore, understanding the generation mechanisms for "spontaneous" or "intrinsic" rotation (i.e., plasma rotation self-consistently generated without external momentum sources) is particularly important in next generation devices.

In recent years, intensive research on intrinsic rotation focused on residual stress which is the non-diffusive and non-convective component of Reynolds stress [10]. The divergence of residual stress is thought to be the origin of the intrinsic rotation. Theoretical works have considered various symmetry breaking mechanisms for the generation of residual stress, such as $\mathbf{E} \times \mathbf{B}$ shear [11, 12], charge separation from polarization drift [13], intensity gradient [14], geometrical up-down asymmetries [15], etc. In addition, a new mechanism named turbulent acceleration of parallel flow in electrostatic turbulence was proposed in Refs. [16, 17]. The turbulent acceleration cannot be written as a divergence of stress, and is an effective volume-force, so it can serve as an important local source/sink of rotation. This is similar to the turbulent momentum source which was proposed in Ref. [18]. One may ask whether the turbulent acceleration contradicts momentum conservation. The answer is no. It is shown that the conserved quantity in electromagnetic turbulence corresponding to axial symmetry is the total canonical momentum density (summing over both species) [19] or the total momentum carried by particles and electromagnetic fields, but not the ion flow velocity [20]. Therefore, turbulent acceleration of parallel flow does not imply that momentum conservation is destroyed.

In previous works, electromagnetic effects on toroidal momentum flux including diffusivity and off-diagonal component were reported [21, 22]. The intrinsic stress force driven by electromagnetic turbulence has also been done [23, 24]. Experimental results also found that the electromagnetic effects on intrinsic rotation may be considerable [25, 26]. Kinetic stress resulting from the correlation between ion pressure and magnetic fluctuations was experimentally identified to drive the intrinsic parallel plasma flow in MST [25]. In DIII-D, the fluid Reynolds stress driven by electrostatic turbulence cannot explain the observed edge rotation profile, and Maxwell stress or kinetic stress could be the remaining candidates for this disagreement [26]. Electromagnetic effects can be particularly important in H-mode pedestal [27]. Electromagnetic ion temperature gradient (ITG) is found to be dominant instability at the pedestal top of DIII-D by gyrokinetic simulation [28]. Therefore, the goal of this work is to investigate the intrinsic rotation drive including both the stress force and the turbulent acceleration in electromagnetic ITG turbulence.

In this work, we theoretically study the electromagnetic effects on intrinsic parallel rotation drive using electromagnetic gyrokinetic equations. Mean parallel flow velocity equation rather than parallel momentum equation is investigated without taking toroidal effects into account. Quasilinear estimates for various stresses including residual stress, kinetic stress and cross Maxwell stress as well as turbulent acceleration driven by electromagnetic ITG turbulence are presented. For the typical electromagnetic ITG turbulence parameters, electromagnetic effects reduce the non-resonant electrostatic stress force and even reverse it, but enhance the resonant stress force. While both the non-resonant and resonant turbulent acceleration terms are enhanced by electromagnetic effects. We also discuss the possible relevance of our results, particularly the electromagnetic turbulent acceleration to the experimental observations.

The remainder of this paper is organized as follows. In Section 2, the evolution equation of mean parallel flow velocity is presented, and quasi-linear estimates for residual stress, kinetic stress, cross Maxwell stress and the turbulent acceleration driven by electromagnetic ITG turbulence are also investigated. Finally, a summary of our work and possible implications for experimental observations are given in Section 3.

## 2. Quasilinear estimates for intrinsic parallel flow drive

The mean parallel flow velocity equation can be derived from the nonlinear electromagnetic gyrokinetic equation [29],

$$\frac{\partial(FB_\parallel^*)}{\partial t} + \nabla \cdot \left(\frac{d\mathbf{R}}{dt} FB_\parallel^*\right) + \frac{\partial}{\partial v_\parallel}\left(\frac{dv_\parallel}{dt} FB_\parallel^*\right) = 0, \tag{1}$$

with gyrocenter equations of motion in the symplectic formulation, i.e., $v_\parallel$

representation

$$\frac{d\mathbf{R}}{dt} = v_\parallel \widehat{\mathbf{b}}^* + \frac{c}{eB} \widehat{\mathbf{b}} \times e\nabla\langle\langle\delta\phi\rangle\rangle, \tag{2}$$

and

$$\frac{dv_\parallel}{dt} = -\frac{e}{m_i}\left(\widehat{\mathbf{b}}^* \cdot \nabla\langle\langle\delta\phi\rangle\rangle + \frac{1}{c}\frac{\partial\langle\langle\delta A_\parallel\rangle\rangle}{\partial t}\right). \tag{3}$$

Here, uniform equilibrium magnetic field $\mathbf{B}$ is assumed. $F = F(R, \mu, v_\parallel, t)$ is the gyrocenter distribution function, $B_\parallel^* = \widehat{\mathbf{b}} \cdot \mathbf{B}^* = B$ is the Jacobian of the transformation from the particle phase space to the gyrocenter phase space with $\mathbf{B}^* = \mathbf{B} + \delta\mathbf{B}_\perp$, $\delta\mathbf{B}_\perp = -\widehat{\mathbf{b}} \times \nabla\delta A_\parallel$. $\widehat{\mathbf{b}}^* = \widehat{\mathbf{b}} + \delta\widehat{\mathbf{B}}_\perp$ with $\delta\widehat{\mathbf{B}}_\perp = \frac{\delta \mathbf{B}_\perp}{B}$. In this paper, the index $\parallel$ refers to components parallel to the equilibrium magnetic field, and the index $\perp$ refers to components perpendicular to the equilibrium magnetic field. Only the shear component of magnetic perturbation, i.e., $\delta A_\parallel$, is considered. $m_i$ is the ion mass, $\delta\phi$ is the electric potential fluctuation, and $\langle\langle\cdots\rangle\rangle$ denotes gyroaveraging. The interested readers can refer to Refs. [16, 17, 20] for the details of calculation. Here, we just directly show the mean field equation for the parallel flow velocity

$$\frac{\partial\langle U_\parallel\rangle}{\partial t} + \nabla \cdot \langle\delta\mathbf{v}_{E\times B,r}\delta U_\parallel\rangle + \frac{1}{m_i n_0}\nabla \cdot \langle\delta\widehat{\mathbf{B}}_r\delta P_\parallel\rangle + \frac{e}{m_i}\nabla \cdot \langle\delta\widehat{\mathbf{B}}_r\delta\phi\rangle$$

$$= \frac{1}{m_i n_0}\left[\langle\delta\hat{n}\widehat{\mathbf{b}} \cdot \nabla\delta P_\parallel\rangle + \langle\delta\hat{n}\delta\widehat{\mathbf{B}}_r\rangle \cdot \nabla P_\parallel\right]. \tag{4}$$

Here, a long wave length limit $k_\perp^2 \rho_i^2 \ll 1$ has been used, with $k_\perp$ being the perpendicular wave number and $\rho_i$ being the ion Larmor radius. $\delta\hat{n} = \frac{\delta n}{n_0}$ is the normalized ion gyrocenter density fluctuation, $\delta\widehat{B}_r = \frac{\delta B_r}{B}$ and $P_\parallel = 2\pi \int dv_\parallel d\mu\, B_\parallel^* v_\parallel^2 F_{i0} = n_0 T_i$ is the equilibrium parallel ion pressure. The kinetic stress $\langle\delta\widehat{\mathbf{B}}_r\delta P_\parallel\rangle$ is somewhat analogous to the kinetic dynamo physics where it is a transport process of electron parallel momentum (current) [30, 31] with $\delta P_\parallel$ being the parallel ion gyrocenter pressure fluctuation whose definition will be presented in the appendix.

In MST, kinetic stress was experimentally identified to drive the intrinsic parallel plasma flow [25]. $\langle\delta\widehat{\boldsymbol{B}}_r\delta\phi\rangle$ is the cross Maxwell stress, which was also presented in Ref [24]. Both kinetic stress and cross Maxwell stress enter the parallel flow equation via their divergence. In this sense, they are surface force, which are similar to the usual Reynolds stress. In contrast, the turbulent acceleration on the right hand side (RHS) of Eq. (4) cannot be recast into a divergence, and is an effective volume-force. The turbulent acceleration driven by parallel gradient of ion pressure fluctuation, $\frac{1}{m_i n_0}\langle\delta\hat{n}\widehat{\boldsymbol{b}}\cdot\nabla\delta P_\parallel\rangle$ was investigated for electrostatic turbulence in Refs. [16, 17], and we denote it by $a_P$ here. The other one is related to the correlation between density and magnetic fluctuation and equilibrium ion pressure gradient, $\frac{1}{m_i n_0}\langle\delta\hat{n}\delta\widehat{\boldsymbol{B}}_r\rangle\cdot\nabla P_\parallel$, and we denote it by $a_M$ here.

Now, we present a quasilinear estimation of the intrinsic flow drive by electromagnetic ITG turbulence in slab geometry including kinetic stress $\langle\delta\widehat{\boldsymbol{B}}_r\delta P_\parallel\rangle$, cross Maxwell stress $\langle\delta\widehat{\boldsymbol{B}}_r\delta\phi\rangle$, residual stress and the turbulent acceleration terms $a_P=\frac{1}{m_i n_0}\langle\delta\hat{n}\widehat{\boldsymbol{b}}\cdot\nabla\delta P_\parallel\rangle$ and $a_M=\frac{1}{m_i n_0}\langle\delta\hat{n}\delta\widehat{\boldsymbol{B}}_r\rangle\cdot\nabla P_\parallel$. We will replace the ion gyrocenter density fluctuation by electron density fluctuation using the quasi-neutrality condition and neglecting the ion polarization density, which is consistent with the long wavelength approximation. This simplification was also used for intrinsic rotation drive by electrostatic ITG turbulence [16]. With assumptions of isothermal electrons along the total magnetic field and collisionless parallel electron force balance for the adiabatic limit $|\omega|\ll k_\parallel v_{the}$ with $k_\parallel v_{the}$ being the electron transit frequency, the generalized electromagnetic adiabatic relation for electrons combined with the electron continuity equation and the Ampère's law yields the normalized density response [32, 33]

$$\delta\hat{n}_k = \delta\hat{\phi}_k + \frac{\omega_{*e}-\omega_k}{\omega_{*e}}\frac{1}{k_\parallel L_n}\delta\hat{A}_{\parallel,k} = (1+\delta_k)\delta\hat{\phi}_k, \tag{5}$$

the normalized magnetic vector potential fluctuation

$$\delta\hat{A}_{\parallel,k} = k_\theta \rho_s c_s \frac{k_\parallel(\omega_{*e}-\omega_k)}{\omega_k(\omega_{*e}-\omega_k)+M}\delta\hat{\phi}_k, \tag{6}$$

and the normalized radial magnetic field fluctuation

$$\delta \hat{B}_{r,k} = i\delta \hat{A}_{\parallel,k} = ik_\theta \rho_s c_s \frac{k_\parallel(\omega_{*e}-\omega_k)}{\omega_k(\omega_{*e}-\omega_k)+M} \delta \hat{\phi}_k. \tag{7}$$

Here, $\omega_k = \omega_r + i\gamma_k$ with $\omega_r$ and $\gamma_k$ being the real frequency and linear growth rate, respectively, and $|\gamma_k| \ll |\omega_r|$. $\delta \hat{\phi}_k = \frac{e\delta \phi_k}{T_e}$ is the normalized electric potential fluctuation, $\delta \hat{A}_{\parallel,k} = \frac{k_\theta \delta A_{\parallel,k}}{B}$ with $k_\theta$ being the poloidal wave number, $M = k_\perp^2 \rho_s^2 k_\parallel^2 v_A^2$ with $k_\parallel$ being the parallel wave number and $v_A = B/\sqrt{4\pi n_0 m_i}$ being the Alfvén velocity, $\omega_{*e} = \frac{k_\theta \rho_s c_s}{L_n}$ is the electron diamagnetic drift frequency with $\rho_s = c_s/\Omega_{ci}$, $c_s = \sqrt{T_e/m_i}$ and $L_n = -(\nabla \ln n_0)^{-1}$ being the density gradient scale length. $\delta_k = -\frac{\omega_k(\omega_{*e}-\omega_k)}{\omega_k(\omega_{*e}-\omega_k)+M} + \frac{\omega_{*e}(\omega_{*e}-\omega_k)}{\omega_k(\omega_{*e}-\omega_k)+M}$ is the electromagnetic contributions to density response with the first term coming from the inductive electric field and the second term coming from the pressure gradient along $\delta \hat{B}_r$. The details can be referred to Ref. [33], but the difference is toroidal effects are not remained here.

The linearized electromagnetic perturbed ion distribution function in Fourier space can be written as

$$\delta f_{ik} = -i \frac{\left\{\tau k_\parallel v_{thi} x_\parallel + \omega_{*e}\left[1+\eta_i \frac{1}{2}(x_\parallel^2+x_\perp^2-3)\right]\right\}\delta \hat{\phi}_k F_{i0} - \left\{\frac{v_{thi} x_\parallel}{L_n}\left[1+\eta_i \frac{1}{2}(x_\parallel^2+x_\perp^2-3)\right]\right\}\delta \hat{A}_{\parallel,k} F_{i0}}{-i(\omega_k - k_\parallel v_\parallel)}. \tag{8}$$

The inverse ion propagator can be written as

$$\frac{1}{-i(\omega_k - k_\parallel v_\parallel)} \approx \frac{i}{\omega_k} + \pi \delta(\omega_r - k_\parallel v_\parallel) = \frac{i}{\omega_k} + \pi \frac{1}{|k_\parallel| v_{thi}} \delta(x_\parallel - \zeta). \tag{9}$$

Here, $\zeta = \frac{\omega_r}{k_\parallel v_{thi}}$, $x_\parallel = v_\parallel/v_{thi}$ with $v_{thi} = \sqrt{T_i/m_i}$, $x_\perp = \sqrt{2\mu B/T_i}$, $\eta_i = L_n/L_{T_i}$ with $L_{T_i} = -(\nabla \ln T_i)^{-1}$ being the ion temperature gradient scale length. $\tau = T_e/T_i$ is the ratio of electron temperature to ion temperature, and $F_{i0}$ is assumed to be Maxwellian. The lowest order of non-resonant part is kept, which is appropriate for drift waves in pedestal region. Then, $\delta f_{ik}$ can be divided into non-resonant and resonant components, i.e., $\delta f_{ik} = \delta f_{ik}^{\text{NR}} + \delta f_{ik}^{\text{Res}}$, with

$$\delta f_{ik}^{\text{NR}} = \frac{1}{\omega_k}\left\{\tau k_\parallel v_{thi} x_\parallel + \omega_{*e}\left[1+\eta_i \frac{1}{2}(x_\parallel^2+x_\perp^2-3)\right]\right\}\delta \hat{\phi}_k F_{i0} - \frac{1}{\omega_k}\left\{\frac{v_{thi} x_\parallel}{L_n}\left[1+\eta_i \frac{1}{2}(x_\parallel^2+x_\perp^2-3)\right]\right\}\delta \hat{A}_{\parallel,k} F_{i0},$$
(10)

and

$$\delta f_{ik}^{\text{Res}} = -i\pi \frac{1}{|k_\parallel|v_{thi}} \delta(x_\parallel - \zeta) F_{i0} \left\{ \left\{ \tau k_\parallel v_{thi} x_\parallel + \omega_{*e} \left[ 1 + \eta_i \frac{1}{2}(x_\parallel^2 + x_\perp^2 - 3) \right] \right\} \delta\hat\phi_k - \right.$$

$$\left. \left\{ \frac{v_{thi}x_\parallel}{L_n} \left[ 1 + \eta_i \frac{1}{2}(x_\parallel^2 + x_\perp^2 - 3) \right] \right\} \delta\hat A_{\parallel,k} \right\}.$$

(11)

Intrinsic rotation drive can thus be also divided into non-resonant drive and resonant drive, respectively. The details of the calculation are presented in the appendix. In the following, we just directly write the expressions.

*2.1 Non-resonant drive.* The non-resonant residual stress can be written as

$$\Pi_{res}^{\text{NR}} = \sum_k \Pi_{ES,k}^{\text{NR}} \left\{ 1 - \frac{K_i}{\tau} \frac{\omega_{*e}[2\omega_r(\omega_{*e}-\omega_r)^2 + M\omega_{*e}]}{[\omega_r(\omega_{*e}-\omega_r) + M]^2} \right\}. \tag{12}$$

Here, $\sum_k \Pi_{ES,k}^{\text{NR}} = \sum_k \rho_s c_s^3 \frac{|\gamma_k|}{\omega_r^2} k_\theta k_\parallel |\delta\hat\phi_k|^2$ is electrostatic non-resonant residual stress, which is in agreement with previous result of Ref. [14], $K_i = (1 + \eta_i)$. The electromagnetic residual stress is driven by equilibrium ion pressure gradient along $\delta\hat B_r$. Kinetic stress can be obtained by combining the ion pressure and magnetic fluctuations and its non-resonant part is as follows,

$$\Pi_{kin}^{\text{NR}} = m_i n_0 \frac{K_i}{\tau} \sum_k \Pi_{ES,k}^{\text{NR}} \frac{M\omega_{*e}(2\omega_r - \omega_{*e})}{[\omega_r(\omega_{*e}-\omega_r) + M]^2}. \tag{13}$$

The cross Maxwell stress is expressed as

$$\Pi_{Max}^{\text{NR}} = \sum_k \Pi_{ES,k}^{\text{NR}} \frac{\omega_r^2[(\omega_{*e}-\omega_r)^2 + M]}{[\omega_r(\omega_{*e}-\omega_r) + M]^2}. \tag{14}$$

For convenience of description, we let $\pm L$ be the scale length of radial variation of the stresses, where $\pm$ corresponds to positive (negative) gradient. Then, the total intrinsic rotation drive due to stress force can be written as

$$\Lambda_{tot}^{\text{NR}} = -\nabla \cdot \Pi_{res}^{\text{NR}} - \frac{1}{m_i n_0} \nabla \cdot \Pi_{kin}^{\text{NR}} - \nabla \cdot \Pi_{Max}^{\text{NR}}$$

$$= \mp \frac{1}{L} \sum_k \Pi_{ES,k}^{\text{NR}} \left\{ 1 + \frac{\omega_r^2[(\omega_{*e}-\omega_r)^2 + M]}{[\omega_r(\omega_{*e}-\omega_r) + M]^2} - \frac{K_i}{\tau} \frac{2\omega_{*e}(\omega_{*e}-\omega_r)}{\omega_r(\omega_{*e}-\omega_r) + M} \right\}. \tag{15}$$

The turbulent acceleration driven by parallel gradient of ion pressure fluctuation can be written as

$$a_p^{\text{NR}} = \frac{1}{L_n} \frac{K_i}{\tau} \sum_k \Pi_{ES,k}^{\text{NR}} \left\{ 1 + \frac{-\omega_r^2(\omega_{*e}-\omega_r)^2 + M(\omega_{*e}-\omega_r)(\omega_{*e}-3\omega_r)}{[\omega_r(\omega_{*e}-\omega_r) + M]^2} \right\}. \tag{16}$$

Here, the first term has the same form as the turbulent acceleration in electrostatic case [16], and the second one comes from the correlation between electromagnetic density response and electrostatic ion pressure fluctuation. The turbulent acceleration driven by equilibrium ion pressure gradient along $\delta \hat{B}_r$ can be obtained by combining magnetic and density fluctuations

$$a_M^{\text{NR}} = -\frac{1}{L_n}\frac{K_i}{\tau}\sum_k \Pi_{ES,k}^{\text{NR}} \frac{M\omega_r^2}{[\omega_r(\omega_{*e}-\omega_r)+M]^2}. \tag{17}$$

Then, the total turbulent acceleration can be written as

$$a_{tot}^{\text{NR}} = \frac{1}{L_n}\frac{K_i}{\tau}\sum_k \Pi_{ES,k}^{\text{NR}} \left\{1 + \frac{-\omega_r^2(\omega_{*e}-\omega_r)^2 + M(\omega_{*e}-\omega_r)(\omega_{*e}-3\omega_r)-M\omega_r^2}{[\omega_r(\omega_{*e}-\omega_r)+M]^2}\right\}. \tag{18}$$

We take the typical parameters from the pedestal top region on DIII-D, where the dominant instability is electromagnetic ITG [28], the safety factor q ≈ 3, $T_e \approx T_i \approx$ 0.64keV, $\tau \approx 1$, $\beta_e = \frac{8\pi n_0 T_e}{B^2} \approx 0.4\%$, $\frac{R_0}{L_n} \approx 2$, $\frac{R_0}{L_{Ti}} \approx 30$, $R_0 = 1.67$m, B = 2T, $m_i = 2m_p$(D plamsa discharge), $k_\parallel \approx \frac{1}{qR_0}$ and $k_\theta \rho_s \approx 0.17$[28], which is consistent with the long wavelength approximation. Based on these parameters, we can assume $\omega_r \approx -2\omega_{*e}$, and estimate the electromagnetic effects on intrinsic flow drive,

$$\Lambda_{tot}^{\text{NR}} = \mp\frac{1}{L}\sum_k \Pi_{ES,k}^{\text{NR}}\left(1 - 4\hat{\beta}_e - 20\hat{\beta}_e + \hat{\beta}_e\right), \tag{19}$$

and

$$a_{tot}^{NR} = 16\frac{1}{L_n}\sum_k \Pi_{ES,k}^{\text{NR}}\left(1 + \frac{21}{4}\hat{\beta}_e - \hat{\beta}_e\right). \tag{20}$$

Here, $\hat{\beta}_e = 4\frac{\omega_{*e}^2}{M} \approx \beta_e \frac{q^2 R_0^2}{L_n^2} = 0.144 \ll 1$ with $k_\perp^2 = 2k_\theta^2$. Therefore, only the lowest order of electromagnetic effects was kept. The second term in the bracket of Eq. (19) is due to electromagnetic residual stress, the third term comes from kinetic stress and the last term results from cross Maxwell stress. Similarly, the second term in the bracket of Eq. (20) is due to the electromagnetic effects on $a_p^{\text{NR}}$, and the third term comes from $a_M^{\text{NR}}$. We find that the main electromagnetic effects on stress force come from electromagnetic residual stress and kinetic stress. The sign of these electromagnetic stresses is opposite to that of electrostatic stress, and can even reverse it for the parameters taken in this work. The electromagnetic part of $a_p^{\text{NR}}$ is larger than $a_M^{\text{NR}}$, and the non-resonant turbulent acceleration is increased by electromagnetic effects. It is noted that $k_\parallel$ symmetry breaking is required for both the non-zero stress force and the turbulent acceleration, which has been discussed intensively in previous works [11-15].

2.2 *Resonant drive.* We can obtain the resonant residual stress

$$\Pi_{res}^{Res} = \sum_k \Pi_{ES,k}^{Res} \left\{ \tau + Y \frac{\omega_{*e}}{\omega_r} - Y \frac{\omega_{*e}(\omega_{*e}-\omega_r)}{\omega_r(\omega_{*e}-\omega_r)+M} \right\}. \tag{21}$$

Here, $\sum_k \Pi_{ES,k}^{Res} = \sum_k \frac{1}{2} \sqrt{\frac{2\pi}{\tau}} \rho_s c_s^2 k_\theta \frac{k_\parallel}{|k_\parallel|} \zeta^2 e^{-\frac{\zeta^2}{2}} |\delta \hat{\phi}_k|^2$, $Y = \left[ 1 + \eta_i \frac{1}{2}(\zeta^2 - 1) \right]$. And the resonant part of kinetic stress is written as

$$\Pi_{kin}^{Res} = -m_i n_0 \sum_k \Pi_{ES,k}^{Res} \left\{ \left( \tau \frac{\omega_r}{\omega_{*e}} + Y \right) \frac{\omega_{*e}(\omega_{*e}-\omega_r)}{\omega_r(\omega_{*e}-\omega_r)+M} - Y \frac{\omega_r}{\omega_{*e}} \left[ \frac{\omega_{*e}(\omega_{*e}-\omega_r)}{\omega_r(\omega_{*e}-\omega_r)+M} \right]^2 \right\}. \tag{22}$$

Obviously, cross Maxwell stress is not involved in resonant effect. Then, the total resonant stress force is given by

$$\Lambda_{tot}^{Res} = -\nabla \cdot \Pi_{res}^{Res} - \frac{1}{m_i n_0} \nabla \cdot \Pi_{kin}^{Res}$$

$$= \mp \frac{1}{L} \sum_k \Pi_{ES,k}^{Res} \left\{ \tau + Y \frac{\omega_{*e}}{\omega_r} - \left( \tau \frac{\omega_r}{\omega_{*e}} + 2Y \right) \frac{\omega_{*e}(\omega_{*e}-\omega_r)}{\omega_r(\omega_{*e}-\omega_r)+M} + Y \frac{\omega_r}{\omega_{*e}} \left[ \frac{\omega_{*e}(\omega_{*e}-\omega_r)}{\omega_r(\omega_{*e}-\omega_r)+M} \right]^2 \right\}. \tag{23}$$

The resonant turbulent acceleration driven by parallel gradient of ion pressure fluctuation can be obtained

$$a_p^{Res} = \frac{1}{L_n} \sum_k \Pi_{ES,k}^{Res} \left\{ \left( \tau \frac{\omega_r}{\omega_{*e}} + Y \right) - \left[ Y \left( 2 \frac{\omega_r}{\omega_{*e}} - 1 \right) - \tau \frac{\omega_r}{\omega_{*e}} \left( 1 - \frac{\omega_r}{\omega_{*e}} \right) \right] \frac{\omega_{*e}(\omega_{*e}-\omega_r)}{\omega_r(\omega_{*e}-\omega_r)+M} - \left( 1 - \frac{\omega_r}{\omega_{*e}} \right) \frac{\omega_r}{\omega_{*e}} Y \left[ \frac{\omega_{*e}(\omega_{*e}-\omega_r)}{\omega_r(\omega_{*e}-\omega_r)+M} \right]^2 \right\}.$$

(24)

Note that the turbulent acceleration driven by equilibrium ion pressure gradient along $\delta \hat{B}_r$ does not have resonant part since a collisionless adiabatic electron fluid model is used.

Using the same parameters as mentioned above, we can obtain $\zeta^2 = \frac{\omega_r^2}{k_\parallel^2 v_{thi}^2} \approx 4 \frac{q^2 R_0^2}{L_n^2} \frac{k_\theta^2 \rho_s^2 c_s^2}{v_{thi}^2} \approx 4$. Then, we can obtain the total resonant stress force

$$\Lambda_{tot}^{Res} \approx \pm \frac{1}{L} \sum_k \Pi_{ES,k}^{Res} \frac{43}{4} \left\{ 1 + \frac{141}{86} \hat{\beta}_e + \frac{3}{2} \hat{\beta}_e \right\}, \tag{25}$$

and the resonant turbulent acceleration

$$a_p^{Res} \approx \frac{1}{L_n} \sum_k \Pi_{ES,k}^{Res} \frac{43}{2} \left\{ 1 + \frac{387}{172} \hat{\beta}_e \right\}. \tag{26}$$

Here, the second term in the bracket of Eq. (25) comes from the electromagnetic resonant residual stress, the third term comes from resonant kinetic stress, and these two terms are comparable. As we can see, different from the non-resonant situation, the electromagnetic effects enhance the resonant stress force as well as the resonant turbulent acceleration. $k_\parallel$ symmetry breaking is also required for both resonant stress force and resonant turbulent acceleration. The results of non-resonant and resonant drive by electromagnetic ITG turbulence are summarized and compared in table 1.

One may wonder whether the resonant drive is considerable or not when it is compared to the non-resonant drive. Here, we make a rough estimation by balancing the total non-resonant stress force and the total resonant stress force

$$\left|\sum_k \Pi_{ES,k}^{NR}\left(1-23\hat{\beta}_e\right)\right| \sim \left|\sum_k \Pi_{ES,k}^{Res} \frac{43}{4}\left(1+\frac{135}{43}\hat{\beta}_e\right)\right|. \tag{27}$$

Then, we can obtain $\left|\frac{\gamma_k}{\omega_r}\right| \approx 1$. In addition, if we balance the total non-resonant turbulent acceleration and the total resonant turbulent acceleration

$$\left|\sum_k \Pi_{ES,k}^{NR} 16\left(1+\frac{17}{4}\hat{\beta}_e\right)\right| \sim \left|\sum_k \Pi_{ES,k}^{Res} \frac{43}{2}\left(1+\frac{387}{172}\hat{\beta}_e\right)\right|, \tag{28}$$

we can also obtain $\left|\frac{\gamma_k}{\omega_r}\right| \approx 0.2$. Therefore, the resonant drive is larger than the non-resonant drive for $|\gamma_k| \ll |\omega_r|$ which is required for quasilinear theory. In a view of the parameter regime, the real frequency of the turbulence $\omega_r \approx -2\omega_{*e}$ is not too big as compared to $k_\parallel v_{thi}$ due to relatively flat density profile in the pedestal top region. Thus, resonant effects can be significant.

**Table 1.** Results of non-resonant and resonant intrinsic flow drive

| Intrinsic flow drive | Non-resonant part | Resonant part |
| --- | --- | --- |
| Residual stress force | $\mp\frac{1}{L}\sum_k \Pi_{ES,k}^{NR}\left(1-4\hat{\beta}_e\right)$ | $\pm\frac{1}{L}\sum_k \Pi_{ES,k}^{Res}\frac{43}{4}\left(1+\frac{141}{86}\hat{\beta}_e\right)$ |
| Kinetic stress force | $\mp\frac{1}{L}\sum_k \Pi_{ES,k}^{NR}\left(-20\hat{\beta}_e\right)$ | $\pm\frac{1}{L}\sum_k \Pi_{ES,k}^{Res}\left(\frac{43}{4}*\frac{3}{2}\hat{\beta}_e\right)$ |
| Cross Maxwell stress force | $\mp\frac{1}{L}\sum_k \Pi_{ES,k}^{NR}\left(\hat{\beta}_e\right)$ | — |
| Total stress force | $\mp\frac{1}{L}\sum_k \Pi_{ES,k}^{NR}\left(1-23\hat{\beta}_e\right)$ | $\pm\frac{1}{L}\sum_k \Pi_{ES,k}^{Res}\frac{43}{4}\left(1+\frac{135}{43}\hat{\beta}_e\right)$ |
| Turbulent acceleration $a_p$ | $\frac{1}{L_n}\sum_k \Pi_{ES,k}^{NR} 16\left(1+\frac{21}{4}\hat{\beta}_e\right)$ | $\frac{1}{L_n}\sum_k \Pi_{ES,k}^{Res}\frac{43}{2}\left(1+\frac{387}{172}\hat{\beta}_e\right)$ |
| Turbulent acceleration $a_M$ | $\frac{1}{L_n}\sum_k \Pi^{NR}_{ES,k} 16(-\hat{\beta}_e)$ | – |
| Total Turbulent acceleration $a_\parallel$ | $\frac{1}{L_n}\sum_k \Pi^{NR}_{ES,k} 16\left(1+\frac{17}{4}\hat{\beta}_e\right)$ | $\frac{1}{L_n}\sum_k \Pi^{Res}_{ES,k} \frac{43}{2}\left(1+\frac{387}{172}\hat{\beta}_e\right)$ |

Note: $\Pi^{NR}_{ES,k} = \rho_s c_s^3 \frac{|\gamma_k|}{\omega_r^2} k_\theta k_\parallel |\delta\hat{\phi}_k|^2$; $\Pi^{Res}_{ES,k} = \frac{1}{2}\sqrt{\frac{2\pi}{\tau}}\rho_s c_s^2 k_\theta \frac{k_\parallel}{|k_\parallel|}\zeta^2 e^{-\frac{\zeta^2}{2}}|\delta\hat{\phi}_k|^2$. These results are obtained in the limits $\omega_r \approx -2\omega_{*e}$, $k_\parallel \approx \frac{1}{qR_0}$, $k_\theta \rho_s \approx 0.17$ and $k_\perp^2 = 2k_\theta^2$ for typical parameters of the pedestal top region on DIII-D [28].

## 3. Summary and discussions

In this work, the electromagnetic effects on intrinsic parallel flow drive are presented. By using electromagnetic gyrokinetic theory, we analytically derive the mean parallel flow velocity equation which includes kinetic stress, cross Maxwell stress as well as the usual parallel Reynolds stress and the parallel turbulent acceleration. Except the turbulent acceleration driven by parallel gradient of ion pressure fluctuation which was proposed for electrostatic turbulence, another new turbulent acceleration driven by the correlation between density and magnetic fluctuations and the equilibrium ion pressure gradient along $\delta\hat{B}_r$ is proposed. Quasilinear estimates of the intrinsic flow drive by electromagnetic ITG turbulence can be divided into non-resonant part and resonant part. Electromagnetic effects reduce the non-resonant electrostatic stress force and even reverse it, but enhance the non-resonant turbulent acceleration. For resonant part, both the resonant stress force and the resonant turbulent acceleration are enhanced by electromagnetic effects. These results in this work are obtained in slab geometry with an approximation of adiabatic electrons. Extending our work into toroidal geometry, such as considering kinetic ballooning mode is ongoing [34].

So far, there have been no experimental observations on turbulent acceleration. However, the kinetic stress associated with density fluctuations $\langle\delta\hat{B}_r\delta\hat{n}\rangle$ has been directly measured in MST [25]. This term can be also measured in tokamaks, if the density and magnetic fluctuations measurements are available. Actually, one of our turbulent acceleration terms, $\frac{1}{m_i n_0}\langle\delta\hat{n}\delta\hat{B}_r\rangle\cdot\nabla P_\parallel$ is also related to the kinetic stress. If the ion pressure profile is given, then the turbulent acceleration can be easily obtained. Taking into account the corresponding turbulent acceleration may be necessary for comprehensive understanding intrinsic rotation. Therefore, we suggest experimentalists to explore the evidence for turbulent acceleration. Moreover, it may be also worth testing the theory of turbulent acceleration from gyrokinetic simulation.

**Acknowledgments**

We thank P. H. Diamond, H. Jhang and W. X. Ding for useful discussions. This work was supported by the Ministry of Science and technology of China, under Contract No. 2013GB112002, and the NSFC Grant Nos. 11305071 and 11675059.

**Appendix. Calculations of intrinsic parallel flow drive**

In this work, $\delta P_\parallel$ is defined as $\delta P_\parallel = \int d^3v m_i v_\parallel^2 \delta f_{ik}$. Firstly, we calculate the non-resonant stress force and turbulent acceleration. By taking the first order and the second order moments of the non-resonant perturbed ion distribution, we can obtain

$$\delta U_{\parallel k}^{\mathrm{NR}} \cong k_\parallel \frac{T_e}{m_i} \frac{1}{\omega_k} \delta\hat{\phi}_k - \frac{T_i}{m_i}(1+\eta_i)\frac{1}{L_n}\frac{1}{\omega_k}\delta\hat{A}_{\parallel,k} ,$$

(A1)

and

$$\delta P_{\parallel k}^{\mathrm{NR}} \cong P_i \frac{1}{\omega_k} \omega_{*e}(1+\eta_i)\delta\hat{\phi}_k ,$$

(A2)

respectively. Then, we can obtain the non-resonant residual stress

$$\Pi_{res}^{\mathrm{NR}} = \Pi_{res,\phi\phi}^{\mathrm{NR}} + \Pi_{res,A\phi}^{\mathrm{NR}} ,$$

(A3)

where $\Pi_{res,\phi\phi}^{\mathrm{NR}} = \sum_k \Pi_{ES,k}^{\mathrm{NR}} = \sum_k \rho_s c_s^3 \frac{|\gamma_k|}{\omega_r^2} k_\theta k_\parallel |\delta\hat{\phi}_k|^2$ is the electrostatic residual stress. The electromagnetic residual stress can be written as

$$\Pi_{res,A\phi}^{\mathrm{NR}} = -\frac{K_i}{\tau}\rho_s c_s^3 \sum_k \frac{k_\theta}{L_n}\left\{\frac{|\gamma_k|}{\omega_r^2}\mathrm{Re}(\delta\hat{A}_{\parallel,k}\delta\hat{\phi}_{-k}) - \frac{1}{\omega_r}\mathrm{Im}(\delta\hat{A}_{\parallel,k}\delta\hat{\phi}_{-k})\right\} ,$$

(A4)

where $\mathrm{Re}(\delta\hat{A}_{\parallel,k}\delta\hat{\phi}_{-k}) = k_\theta k_\parallel \rho_s c_s \frac{(\omega_{*e}-\omega_r)}{\omega_r(\omega_{*e}-\omega_r)+M}|\delta\hat{\phi}_k|^2$, and $\mathrm{Im}(\delta\hat{A}_{\parallel,k}\delta\hat{\phi}_{-k}) =$

$-k_\theta k_\| \rho_s c_s \frac{|\gamma_k|[(\omega_{*e}-\omega_r)^2+M]}{[\omega_r(\omega_{*e}-\omega_r)+M]^2}|\delta\hat{\phi}_k|^2$. So the total non-resonant residual stress is

$$\Pi_{res}^{NR} = \sum_k \Pi_{ES,k}^{NR}\left\{1 - \frac{K_i}{\tau}\frac{\omega_{*e}[2\omega_r(\omega_{*e}-\omega_r)+M\omega_{*e}]}{[\omega_r(\omega_{*e}-\omega_r)+M]^2}\right\}$$

(A5)

In a similar way, the non-resonant kinetic stress and cross Maxwell stress can be expressed as

$$\Pi_{kin}^{NR} = \langle \delta\hat{B}_r \delta P_\|^{NR}\rangle$$

$$= -m_i n_0 \frac{K_i}{\tau}\rho_s c_s^3 \sum_k \frac{k_\theta}{L_n}\left\{\frac{|\gamma_k|}{\omega_r^2}\text{Re}(\delta\hat{A}_{\|,k}\delta\hat{\phi}_{-k}) + \frac{1}{\omega_r}\text{Im}(\delta\hat{A}_{\|,k}\delta\hat{\phi}_{-k})\right\}$$

$$= m_i n_0 \frac{K_i}{\tau}\sum_k \Pi_{ES,k}^{NR}\frac{M\omega_{*e}(2\omega_r-\omega_{*e})}{[\omega_r(\omega_{*e}-\omega_r)+M]^2}, \tag{A6}$$

and

$$\Pi_{Max}^{NR} = \frac{e}{m_i}\langle\delta\hat{B}_r\delta\phi\rangle$$

$$= -c_s^2 \sum_k \text{Im}(\delta\hat{A}_{\|,k}\delta\hat{\phi}_{-k})$$

$$= \sum_k \Pi_{ES,k}^{NR}\frac{\omega_r^2[(\omega_{*e}-\omega_r)^2+M]}{[\omega_r(\omega_{*e}-\omega_r)+M]^2}. \tag{A7}$$

Then, the total non-resonant stress force is given by

$$\Lambda_{tot}^{NR} = -\nabla\cdot\Pi_{res}^{NR} - \frac{1}{m_i n_0}\nabla\cdot\Pi_{kin}^{NR} - \nabla\cdot\Pi_{Max}^{NR}$$

$$= \mp\frac{1}{L}\sum_k \Pi_{ES,k}^{NR}\left\{1 + \frac{\omega_r^2[(\omega_{*e}-\omega_r)^2+M]}{[\omega_r(\omega_{*e}-\omega_r)+M]^2} - \frac{K_i}{\tau}\frac{2\omega_{*e}(\omega_{*e}-\omega_r)}{\omega_r(\omega_{*e}-\omega_r)+M}\right\}. \tag{A8}$$

The non-resonant turbulent acceleration terms can be also written as follows

$$a_p^{NR} = \frac{1}{m_i n_0}\langle\delta\hat{n}\hat{\boldsymbol{b}}\cdot\nabla\delta P_\|^{NR}\rangle$$

$$= a_{p,\phi\phi}^{NR} + a_{p,A\phi}^{NR}$$

$$= \frac{1}{L_n}\frac{K_i}{\tau}\sum_k \Pi_{ES,k}^{NR}\left\{1 + \frac{-\omega_r^2(\omega_{*e}-\omega_r)^2+M(\omega_{*e}-\omega_r)(\omega_{*e}-3\omega_r)}{[\omega_r(\omega_{*e}-\omega_r)+M]^2}\right\}, \tag{A9}$$

where $a_{p,\phi\phi}^{NR} = \frac{1}{L_n}\frac{K_i}{\tau}\sum_k \Pi_{ES,k}^{NR}$ is the electrostatic turbulent acceleration, and the electromagnetic contribution is $a_{p,A\phi}^{NR} = \frac{K_i}{\tau}\rho_s c_s^3 \frac{1}{L_n}\sum_k \frac{k_\theta}{L_n}\left\{\frac{|\gamma_k|}{\omega_r^2}\left(1 - \frac{2\omega_r}{\omega_{*e}}\right)\text{Re}(\delta\hat{A}_{\|,k}\delta\hat{\phi}_{-k}) + \frac{1}{\omega_r}\left(1 - \frac{\omega_r}{\omega_{*e}}\right)\text{Im}(\delta\hat{A}_{\|,k}\delta\hat{\phi}_{-k})\right\}$, which results from the electromagnetic density response. The non-resonant turbulent acceleration driven by equilibrium ion pressure along $\delta\hat{\boldsymbol{B}}_r$ can be given as

$$a_M^{NR} = \frac{1}{m_i n_o}\langle\delta\hat{n}\delta\hat{\boldsymbol{B}}_r\rangle\bullet\nabla P_\|$$

$$= a_{M,A\phi}^{NR} + a_{M,AA}^{NR}$$

$$= -\frac{1}{L_n}\frac{K_i}{\tau}\sum_k \Pi_{ES,k}^{NR} \frac{M\omega_r^2}{[\omega_r(\omega_{*e}-\omega_r)+M]^2}, \tag{A10}$$

where $a_{M,A\phi}^{NR} = \frac{K_i}{\tau}c_s^2\frac{1}{L_n}\sum_k \text{Im}(\delta\hat{A}_{\parallel,k}\delta\hat{\phi}_{-k})$, and $a_{M,AA}^{NR} = \frac{K_i}{\tau}c_s^2\frac{1}{L_n}\sum_k \frac{|\gamma_k|}{\omega_{*e}}\frac{1}{k_\parallel L_n}|\delta\hat{A}_{\parallel,k}|^2$,

with $|\delta\hat{A}_{\parallel,k}|^2 = k_\theta^2\rho_s^2 c_s^2 \frac{k_\parallel^2(\omega_{*e}-\omega_r)^2}{[\omega_r(\omega_{*e}-\omega_r)+M]^2}|\delta\hat{\phi}_k|^2$. So the total non-resonant turbulent acceleration can be written as

$$a_{tot}^{NR} = \frac{1}{L_n}\frac{K_i}{\tau}\sum_k \Pi_{ES,k}^{NR}\left\{1 + \frac{-\omega_r^2(\omega_{*e}-\omega_r)^2 + M(\omega_{*e}-\omega_r)(\omega_{*e}-3\omega_r) - M\omega_r^2}{[\omega_r(\omega_{*e}-\omega_r)+M]^2}\right\}. \tag{A11}$$

Next, we calculate the resonant stress force and resonant turbulent acceleration. The resonant part of the first order and the second order of moments can be written as

$$\delta U_{\parallel k}^{Res} = c_s(iA_\phi \delta\hat{\phi}_k + iA_A \delta\hat{A}_{\parallel,k}), \tag{A12}$$

and

$$\delta P_{\parallel k}^{Res} = P_i(iB_\phi \delta\hat{\phi}_k + iB_A \delta\hat{A}_{\parallel,k}), \tag{A13}$$

where $A_\phi = -\frac{1}{2}\sqrt{\frac{2\pi}{\tau}}\frac{k_\parallel}{|k_\parallel|}\zeta^2 e^{-\frac{\zeta^2}{2}}\left(\tau + Y\frac{\omega_{*e}}{\omega_r}\right)$, $A_A = \frac{1}{2}\sqrt{\frac{2\pi}{\tau}}\frac{1}{|k_\parallel|L_n}\zeta^2 e^{-\frac{\zeta^2}{2}}Y$, $B_\phi = -\frac{\sqrt{2\pi}}{2}\frac{k_\parallel}{|k_\parallel|}\zeta^3 e^{-\frac{\zeta^2}{2}}\left(\tau + Y\frac{\omega_{*e}}{\omega_r}\right)$, $B_A = \frac{\sqrt{2\pi}}{2}\frac{1}{|k_\parallel|L_n}\zeta^3 e^{-\frac{\zeta^2}{2}}Y$, with $Y = \left[1 + \eta_i\frac{1}{2}(\zeta^2 - 1)\right]$.

Then, we can obtain the resonant residual stress

$$\Pi_{res}^{Res} = \Pi_{res,\phi\phi}^{Res} + \Pi_{res,A\phi}^{Res}$$

$$= \sum_k \Pi_{ES,k}^{Res}\left\{\tau + Y\frac{\omega_{*e}}{\omega_r} - Y\frac{\omega_{*e}(\omega_{*e}-\omega_r)}{\omega_r(\omega_{*e}-\omega_r)+M}\right\}, \tag{A14}$$

where $\Pi_{ES,k}^{Res} = \frac{1}{2}\sqrt{\frac{2\pi}{\tau}}\rho_s c_s^2 k_\theta \frac{k_\parallel}{|k_\parallel|}\zeta^2 e^{-\frac{\zeta^2}{2}}|\delta\hat{\phi}_k|^2$, $\Pi_{res,\phi\phi}^{Res} = -\sum_k k_\theta \rho_s c_s^2 A_\phi |\delta\hat{\phi}_k|^2 = \sum_k \Pi_{ES,k}^{Res}\left(\tau + Y\frac{\omega_{*e}}{\omega_r}\right)$ is the electrostatic resonant residual stress, and $\Pi_{res,A\phi}^{Res} = -\sum_k k_\theta \rho_s c_s^2 A_A \text{Re}(\delta\hat{A}_{\parallel,k}\delta\hat{\phi}_{-k}) = -\sum_k \Pi_{ES,k}^{Res} Y \frac{\omega_{*e}(\omega_{*e}-\omega_r)}{\omega_r(\omega_{*e}-\omega_r)+M}$ is the electromagnetic residual stress. The resonant kinetic stress can be written as

$$\Pi_{kin}^{Res} = \Pi_{kin,A\phi}^{Res} + \Pi_{kin,AA}^{Res}$$

$$= -m_i n_0 \sum_k \Pi_{ES,k}^{Res}\left\{\left(\tau\frac{\omega_r}{\omega_{*e}}+Y\right)\frac{\omega_{*e}(\omega_{*e}-\omega_r)}{\omega_r(\omega_{*e}-\omega_r)+M} - Y\frac{\omega_r}{\omega_{*e}}\left[\frac{\omega_{*e}(\omega_{*e}-\omega_r)}{\omega_r(\omega_{*e}-\omega_r)+M}\right]^2\right\}, \tag{A15}$$

where

$$\Pi_{kin,A\phi}^{Res} = n_0 T_i \sum_k B_\phi \text{Re}(\delta\hat{A}_{\parallel,k}\delta\hat{\phi}_{-k}) = -n_0 T_i \frac{\sqrt{2\pi}}{2}\sum_k \zeta^2 e^{-\frac{\zeta^2}{2}}\frac{k_\parallel}{|k_\parallel|}\left(\tau + Y\frac{\omega_{*e}}{\omega_r}\right)\zeta \text{Re}(\delta\hat{A}_{\parallel,k}\delta\hat{\phi}_{-k}) = -m_i n_0 \sum_k \Pi_{ES,k}^{Res}\left(\tau + Y\frac{\omega_{*e}}{\omega_r}\right)\frac{\omega_r(\omega_{*e}-\omega_r)}{\omega_r(\omega_{*e}-\omega_r)+M},$$

and
$$\Pi_{kin,AA}^{Res} = n_0 T_i \sum_k B_A \left|\delta \hat{A}_{\parallel,k}\right|^2 = n_0 T_i \frac{\sqrt{2\pi}}{2} \sum_k \zeta^2 e^{-\frac{\zeta^2}{2}} Y \frac{1}{|k_\parallel| L_n} \zeta \left|\delta \hat{A}_{\parallel,k}\right|^2 =$$
$m_i n_0 \sum_k \Pi_{ES,k}^{Res} Y \frac{\omega_r}{\omega_{*e}} \left[\frac{\omega_{*e}(\omega_{*e}-\omega_r)}{\omega_r(\omega_{*e}-\omega_r)+M}\right]^2$. Then, the total resonant stress force can be written as

$$\Lambda_{tot}^{Res} = -\nabla \cdot \Pi_{res}^{Res} - \frac{1}{m_i n_0} \nabla \cdot \Pi_{kin}^{Res}$$
$$= \mp \frac{1}{L} \sum_k \Pi_{ES,k}^{Res} \left\{\tau + Y \frac{\omega_{*e}}{\omega_r} - \left(\tau \frac{\omega_r}{\omega_{*e}} + 2Y\right) \frac{\omega_{*e}(\omega_{*e}-\omega_r)}{\omega_r(\omega_{*e}-\omega_r)+M} + Y \frac{\omega_r}{\omega_{*e}} \left[\frac{\omega_{*e}(\omega_{*e}-\omega_r)}{\omega_r(\omega_{*e}-\omega_r)+M}\right]^2\right\}.$$
(A16)

The resonant turbulent acceleration driven by parallel gradient of ion pressure fluctuation can be obtained

$$a_p^{Res} = \frac{1}{m_i n_0} \langle \delta \hat{n} \hat{\boldsymbol{b}} \cdot \nabla \delta P_\parallel^{Res}\rangle$$
$$= a_{p,\phi\phi}^{Res} + a_{p,A\phi}^{Res} + a_{p,AA}^{Res}$$
$$= \frac{1}{L_n} \sum_k \Pi_{ES,k}^{Res} \left\{\left(\tau \frac{\omega_r}{\omega_{*e}} + Y\right) - \left[Y\left(2 \frac{\omega_r}{\omega_{*e}} - 1\right) - \tau \frac{\omega_r}{\omega_{*e}}\left(1 - \frac{\omega_r}{\omega_{*e}}\right)\right] \frac{\omega_{*e}(\omega_{*e}-\omega_r)}{\omega_r(\omega_{*e}-\omega_r)+M} - \left(1 - \frac{\omega_r}{\omega_{*e}}\right) \frac{\omega_r}{\omega_{*e}} Y \left[\frac{\omega_{*e}(\omega_{*e}-\omega_r)}{\omega_r(\omega_{*e}-\omega_r)+M}\right]^2\right\},$$
(A17)

where $a_{p,\phi\phi}^{Res} = -v_{thi}^2 \sum_k k_\parallel B_\phi |\delta \hat{\phi}_k|^2 = \frac{1}{L_n} \sum_k \Pi_{ES,k}^{Res} \left(\tau \frac{\omega_r}{\omega_{*e}} + Y\right)$ is the electrostatic resonant acceleration,

$$a_{p,A\phi}^{Res} = -v_{thi}^2 \sum_k \left(B_\phi \frac{\omega_{*e}-\omega_r}{\omega_{*e}} \frac{1}{L_n} + k_\parallel B_A\right) \text{Re}\left(\delta \hat{A}_{\parallel,k} \delta \hat{\phi}_{-k}\right)$$
$$= -\frac{1}{L_n} \frac{\sqrt{2\pi}}{2} c_s^2 \frac{1}{\tau} \sum_k \frac{k_\parallel}{|k_\parallel|} \zeta^2 e^{-\frac{\zeta^2}{2}} \left[Y\left(2 - \frac{\omega_{*e}}{\omega_r}\right) - \tau\left(1 - \frac{\omega_r}{\omega_{*e}}\right)\right] \zeta \text{Re}\left(\delta \hat{A}_{\parallel,k} \delta \hat{\phi}_{-k}\right)$$
$$= -\frac{1}{L_n} \sum_k \Pi_{ES,k}^{Res} \left[Y\left(2 \frac{\omega_r}{\omega_{*e}} - 1\right) - \tau \frac{\omega_r}{\omega_{*e}}\left(1 - \frac{\omega_r}{\omega_{*e}}\right)\right] \frac{\omega_{*e}(\omega_{*e}-\omega_r)}{\omega_r(\omega_{*e}-\omega_r)+M},$$
(A18)

and

$$a_{p,AA}^{Res} = -v_{thi}^2 \sum_k B_A \frac{\omega_{*e}-\omega_r}{\omega_{*e}} \frac{1}{L_n} \left|\delta \hat{A}_{\parallel,k}\right|^2$$
$$= -\frac{1}{L_n} \frac{\sqrt{2\pi}}{2} c_s^2 \frac{1}{\tau} \sum_k \zeta^2 e^{-\frac{\zeta^2}{2}} Y\left(1 - \frac{\omega_r}{\omega_{*e}}\right) \frac{1}{|k_\parallel| L_n} \zeta \left|\delta \hat{A}_{\parallel,k}\right|^2$$
$$= -\frac{1}{L_n} \sum_k \Pi_{ES,k}^{Res} \left(1 - \frac{\omega_r}{\omega_{*e}}\right) \frac{\omega_r}{\omega_{*e}} Y \left[\frac{\omega_{*e}(\omega_{*e}-\omega_r)}{\omega_r(\omega_{*e}-\omega_r)+M}\right]^2$$
(A19)

are the electromagnetic resonant acceleration terms. The turbulent acceleration driven by equilibrium ion pressure gradient along $\delta\widehat{\boldsymbol{B}}_r$ does not have resonant part since a collisionless adiabatic electron fluid model is used.